\pdfoutput=1

\documentclass[aps,pra, amsmath,superscriptaddress,twocolumn, 10pt]{revtex4-1}
\usepackage{siunitx}
\usepackage[pdftex]{graphicx}
\usepackage{bm}
\usepackage{setspace}

\usepackage{graphicx}
\usepackage{bm}        
\usepackage{textcomp}
\usepackage{gensymb}
\usepackage{amssymb}

\newcommand*{\citen}[1]{%
  \begingroup
    \romannumeral-`\x 
    \setcitestyle{numbers}%
    \cite{#1}%
  \endgroup
}

\begin{document}
\setstretch{1.05}

\title[Width-dependent Photoluminescence and Raman from Monolayer MoS$_2$ Nanoribbons]{Width-dependent Photoluminescence and Anisotropic Raman Spectroscopy from Monolayer MoS$_2$ Nanoribbons}

\author{Guohua Wei}
\affiliation{Applied Physics Program, Northwestern University, 2145 Sheridan Road, Evanston, IL 60208, USA}
\author{Erik J. Lenferink}
\affiliation{Department of Physics and Astronomy, Northwestern University, 2145 Sheridan Road, Evanston, IL 60208, USA}
\author{David A. Czaplewski}
\affiliation{Center for Nanoscale Materials, Argonne National Laboratory, 9700 S Cass Avenue, Argonne, IL 60439, USA}
\author{Nathaniel P. Stern}\email{n-stern@northwestern.edu}
\affiliation{Department of Physics and Astronomy, Northwestern University, 2145 Sheridan Road, Evanston, IL 60208, USA}

\begin{abstract}
\vspace{1em}
Single layers of transition metal dichalcogenides such as MoS$_2$ are direct bandgap semiconductors with optical and electronic properties distinct from multilayers due to strong vertical confinement. Despite the fundamental monolayer limit of thickness, the electronic structure of isolated layers can be further tailored with lateral degrees of freedom in nanostructures such as quantum dots or nanoribbons. Although one-dimensionally confined monolayer semiconductors are predicted to have interesting size- and edge-dependent properties useful for spintronics applications, experiments on the opto-electronic features of monolayer transition metal dichalcogenide nanoribbons is limited. We use nanolithography to create monolayer MoS$_2$ nanoribbons with lateral sizes down to 20~nm. The Raman spectra show polarization anisotropy and size-dependent intensity. The nanoribbons prepared with this technique show reduced susceptibility to edge defects and emit photoluminescence with size-dependent energy that can be understood from a phenomenological model. Fabrication of monolayer nanoribbons with strong exciton emission can facilitate exploration of low-dimensional opto-electronic devices with controllable properties.
\vspace{2em}
\end{abstract}

\maketitle
\vspace{2em}

Atomically thin two dimensional (2D) transition metal dichalcogenides (TMDs), such as monolayer (ML) MoS$_2$, have attracted a lot of attention over the last several years due to their rich physics and potential device applications. In addition to utility in traditional semiconductor devices such as transistors and photodetectors, single layer TMDs have more exotic carrier properties as a result of their crystal symmetry and strong spin-orbit interaction such as spin and valley locking~\cite{Xiao2012,Xu2014}, creating potential for harnessing new phenomena in nano-scale opto-electronics. As with traditional semiconductors, engineering confinement in monolayer nanostructures provides an effective way of tuning electric and optical properties, demonstrated in graphene quantum dots and nanoribbons (NRs)~\cite{Trauzettel2007,Libisch2009,han2007energy,brey2006electronic}.
In addition to size-dependent energy levels, laterally-confined monolayers exhibit edge dependent electrical and magnetic properties as observed in graphene NRs~\cite{dutta2010novel,brey2006electronic} and predicted for ML TMD NRs~\cite{pan2012edge,li2008mos2,ataca2010mechanical,dolui2013electric}. For example, helical edge modes in a 2D topological insulating NR brought in proximity to superconductors can generate Majorana fermions at the ribbon ends~\cite{klinovaja2013spintronics,mourik2012signatures}. The possibility of novel low-dimensional transport in edges and interfaces of ML TMDs, such as conducting interface charge accumulation~\cite{Wu2016edge, Jia2017} or topological insulator edge states in ML WTe$_2$~\cite{ fei2017edge,jia2017direct, tang2017quantum}, makes confined ML NRs interesting for controlling optical, electronic, and spintronic properties in nanomaterials.

Despite this potential for low-dimensional opto-electronics, experimental work on TMD NRs has so far been limited. Bottom-up synthesis techniques show promise but they have so far provided limited control of size and layer number.~\cite{poh2017large}. Transmission electron microscopy tools can combine patterning of nanometer-scale ribbons with in-situ atomic-resolution characterization, allowing observation of vacancy migration along the edges in WS$_{2}$ NRs~\cite{liu2011identification}. This method is not suitable for optical and electric transport studies, however. Recently, top-down patterning using helium ion milling was used to create ML MoS$_{2}$ NRs, which exhibited anisotropic Raman spectra and low frequency Raman edge modes that are absent in the pristine MLs~\cite{wu2016monolayer}. So far, size-tunable excitonic behavior due to lateral confinement, observed in graphene NRs~\cite{han2007energy}, are not reported in TMD NRs. In the previous work, the lack of reported size-dependent excitonic emission is possibly due to creation of high-density defects during the helium ion milling process, which can also cause the low frequency Raman modes such as that observed in monolayer WS$_2$ and WSe$_2$ nanocrystals~\cite{shi2016raman}.  A nondestructive fabrication technique for patterning ML naoribbons with controllable size is needed for excitonic studies.

Here, we use electron beam (e-beam) lithography with reactive ion etching (RIE) to pattern size-tunable ML MoS$_2$ NRs with suitable optical properties. Anisotropic Raman spectra are studied and size-tunable exciton photoluminescence (PL) is observed, which can be explained with a non-equilibrium model. The NRs created with this process are also found to be suitable for electronic transport devices.

\raggedbottom

\begin{figure*}[thb!]
\includegraphics[scale=1]{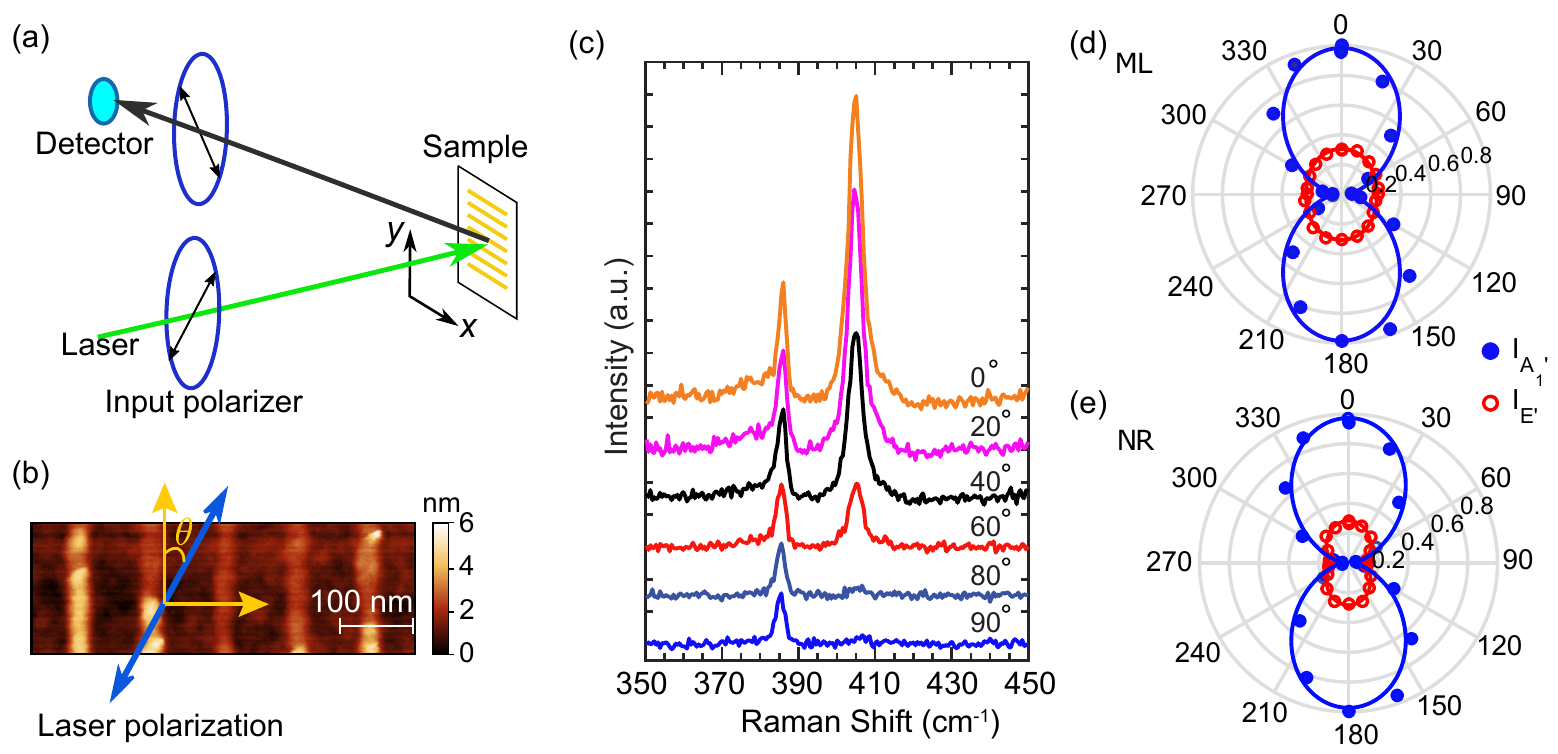}
\caption[Polarization dependent Raman spectra of MoS$_{2}$]{Polarization resolved Raman spectroscopy of MoS$_2$ NRs. (a) Illustration of the geometry for polarized Raman spectroscopy. Polarization of excitation laser and scattered light are independently controlled by the polarizer and analyzer. The incident and collection are both in normal direction to the sample. (b) AFM image of NRs show ribbon width of about 25~nm, with the angle $\theta$ defined between incident laser polarization and the ribbon orientation. (c) Raman spectra of NRs for incident polarization from 0$\degree$ to 90$\degree$. The spectra are shifted in the vertical direction for clarity. (d, e) Polar plot for the integrated Raman intensities of the MoS$_{2}$ ML (d) and NRs (e). The integration is from 370~cm$^{-1}$ to 393~cm$^{-1}$ for the $E^{\prime}$ mode and 393~cm$^{-1}$ to 420~cm$^{-1}$ for the $A_{1}^{\prime}$ mode.}
\label{fig:RamanPolarization1}
\end{figure*}

\section{Sample Preparation and Fabrication}

Patterning of lateral confinement follows similar procedures used for MoS$_2$ nanodots~\cite{Wei2017}. ML MoS$_2$ is obtained through mechanical exfoliation with scotch tape and dry transferred onto a SiO$_2$/Si wafer. A high resolution e-beam lithography system operating at 100~kV (JEOL 9300FS) is used for patterning. The exfoliated ML flake is coated with a positive resist (GL2000) with a thickness over 40~nm. After e-beam exposure and cold development (-5~\degree C), the resist nanowire array is formed on the flake.  A reactive ion etch (RIE) is performed following the e-beam to transfer the nanoribbon pattern to ML MoS$_2$ with desired geometry. Height profiles from atomic force microscopy (AFM) are used for NR width characterization. Details of the fabrication process are in the supplementary materials.

\section{Raman Spectroscopy of Monolayer MoS$_{2}$ Nanoribbons}

To characterize the structural anisotropy of the ML NRs we utilize polarized Raman spectroscopy, a technique commonly used to probe fundamental phonon modes~\cite{lee2010anomalous,molina2011phonons} and crystal symmetries. Since Raman spectroscopy is sensitive to structural changes, it is often applied to study layer- and strain-dependent properties of 2D materials~\cite{zhang2015phonon,saito2016raman,mignuzzi2015effect,conley2013bandgap} and to study nanostructures such as nanocrystals~\cite{faraci2006modified,faraci2009quantum,swamy2005finite}, graphene chiral edge states~\cite{you2008edge}, and graphene and MoS$_2$ NRs~\cite{ryu2011raman,bischoff2011raman, wu2016monolayer}. With polarized Raman spectroscopy, the intensity of scattered light is measured as a function of its incident polarization angle.

ML MoS$_{2}$ has two active Raman modes, $E^{\prime}$ and $A_{1}^{\prime}$, as a result of its D$_{3h}$ crystal symmetry~\cite{saito2016raman,zhang2015phonon}. The $E^{\prime}$ mode corresponds to the in-plane vibration of the two sulphur layers of a ML MoS$_2$ and the $A_{1}^{\prime}$ mode to the out-of-plane vibration of the two sulphur layers. They have different polarization-dependent behaviors in the backscattering measurement scheme illustrated in Fig.~\ref{fig:RamanPolarization1}(a). The $E^{\prime}$ mode has two components with polarization parallel and perpendicular to the incident laser polarization respectively, while the $A_{1}^{\prime}$ mode only has a component with polarization parallel to the incident~\cite{saito2016raman,zhang2015phonon}. By comparing the polarization dependence of the intensity of these two modes between ML MoS$_2$ NRs and unpatterned MLs, the effects of anisotropic lateral confinement in NRs may be measured.

Figure~\ref{fig:RamanPolarization1}(b-c) shows polarization resolved Raman spectra for ML MoS$_{2}$ NRs with a measured ribbon width of about 25~nm, as shown by the AFM image in Fig.~\ref{fig:RamanPolarization1}(b). The analyzer polarization direction is parallel to the ribbons ($x$ direction), and the angle for the incident light polarization is defined as in Fig.~\ref{fig:RamanPolarization1}(b). Figure~\ref{fig:RamanPolarization1}(c) shows the NR Raman spectra with $\theta$ from 0$\degree$ to 90$\degree$. Angle-dependence for both Raman modes is evident. Figure~\ref{fig:RamanPolarization1}(d, e) show polar plots of the integrated intensities of Raman modes for ML and NRs. Due to the crystal symmetry in ML MoS$_2$, the $E^{\prime}$ mode is almost angle independent while the $A_{1}^{\prime}$ mode is $\theta$-dependent. In contrast to unpatterned ML MoS$_{2}$, the $E^{\prime}$ mode in the NRs shows stronger polarization dependence. The intensities for both modes reach maximum (minimum) when incident laser polarization is parallel (perpendicular) to the ribbons. The Raman anisotropy is a clear indication of the symmetry breaking in the ribbon nanostructures. Similar anisotropic Raman spectra are also reported in Ref.~\citen{wu2016monolayer} for monolayer MoS$_2$ created by helium ion milling, however, the maximum (minimum) for both modes are at 90$\degree$ (0\degree) instead of 0\degree. Our results are confirmed with two different Raman setups and are consistent with other literature~\cite{saito2016raman,shi2016raman}.

\begin{figure}[tbp!]
\includegraphics[scale=1]{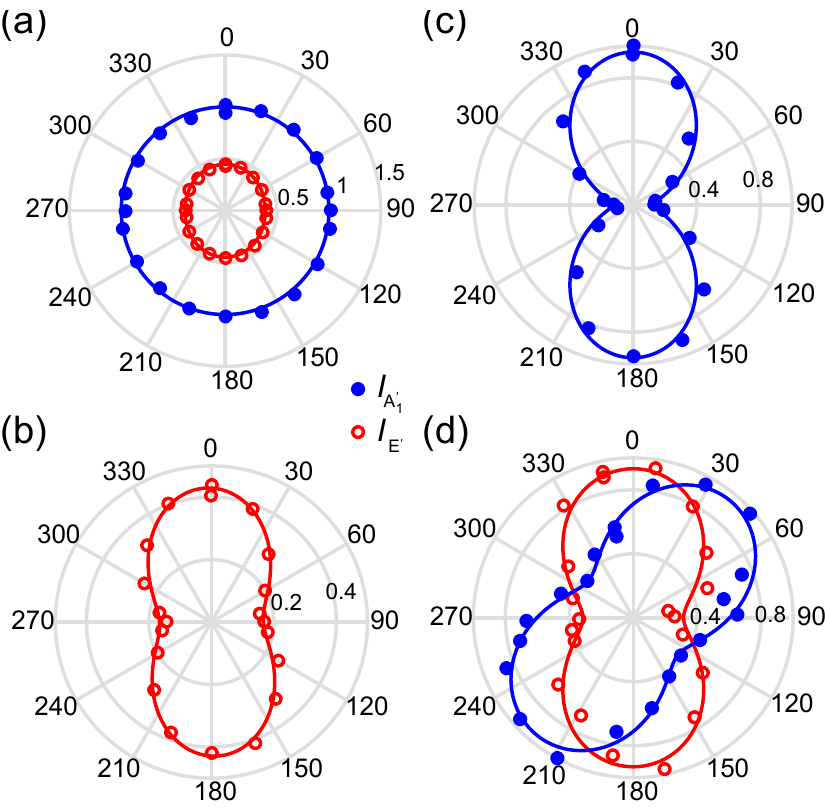}
\caption{Polarized Raman spectra of ML MoS$_2$ and NRs. (a) Polar plot of Raman intensities of ML MoS$_2$ collected without analyzer. (b) $E^{\prime}$ mode of the NRs Raman spectra collected without analyzer. (c) $A_{1}^{\prime}$ mode of the NRs measured without analyzer. (d) Raman intensities of the NRs collected with analyzer perpendicular to the ribbon orientation (analyzer along $y$-axis).}
\label{fig:RamanPol2}
\end{figure}

To further elucidate the anisotropy, Raman spectra are collected without the analyzer. Figure~\ref{fig:RamanPol2}(a) shows the Raman spectra of ML MoS$_{2}$; the intensities $I(A_{1}^{\prime})$ and $I(E^{\prime})$ are nearly isotropic as expected. In contrast, the  $I(E^{\prime})$ and  $I(A_{1}^{\prime})$ for NRs as shown in Fig.~\ref{fig:RamanPol2}(b, c) show strong polarization dependence, clearly revealing the anisotropic Raman behavior. More interestingly, when the analyzer is perpendicular to the NRs, the $A_{1}^{\prime}$ mode is polarized around $45\degree$, as shown in Fig.~\ref{fig:RamanPol2}(d). This indicates that the $A_{1}^{\prime}$ mode is no longer polarized along with the excitation laser but has other components, clearly been modified by the NR structure. The measurement with various polarization configurations revealed the anisotropic Raman spectra in the ML NRs.

\begin{figure*}[bht]
\includegraphics[]{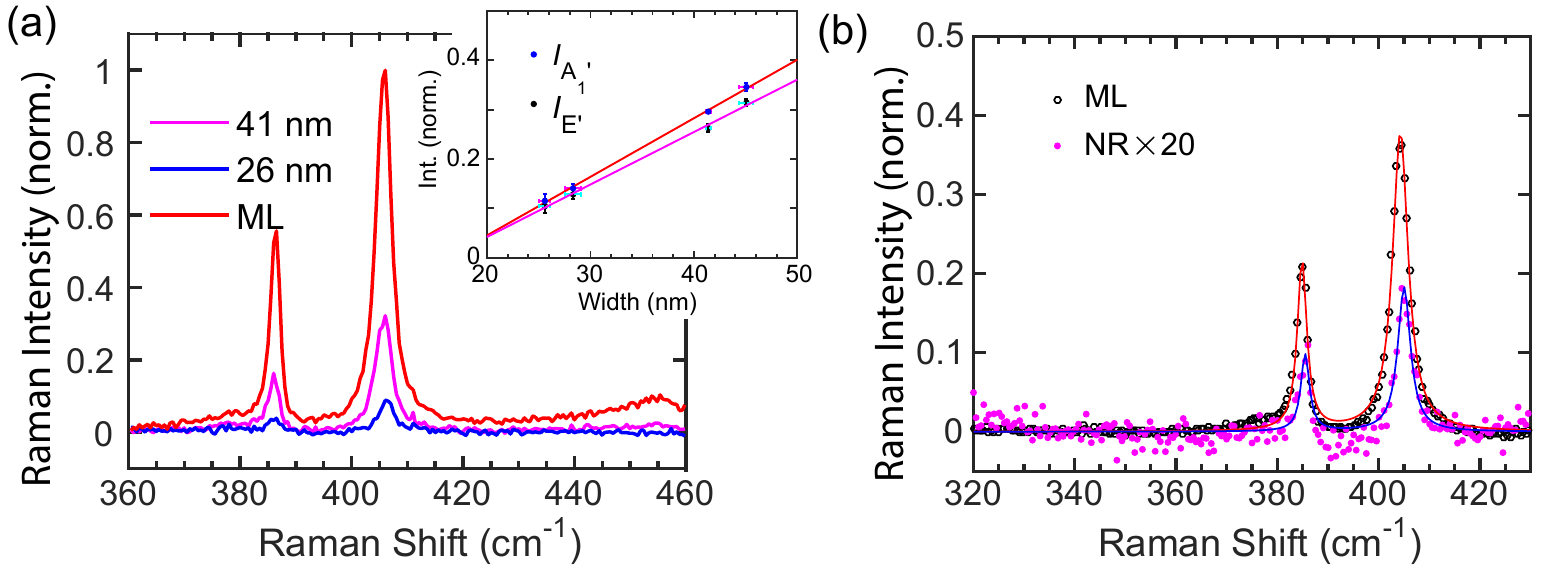}
\caption[Ribbon width-dependent Raman spectroscopy]{Size-dependent Raman spectroscopy. (a) Raman spectra of ML MoS$_{2}$ and NRs with different widths. Insert shows the integrated Raman intensities versus ribbon width. The Raman intensities of NRs are normalized to that of the ML. (b) Raman spectrum of ML MoS$_{2}$ NRs with width of about 15~nm. The Raman intensities are normalized to the silicon line from the substrate, and the spectrum for NRs is magnified by 20 times for comparison.}
\label{fig:RamanWidth}
\end{figure*}

\section{Size-dependence of Raman Spectra}

Figure~\ref{fig:RamanWidth}(a) shows the Raman spectra for NRs with different ribbon widths. The Raman intensity decreases as the ribbon becomes narrower for both modes, as expected.  The integrated Raman intensities of the two modes both follow linear relations to the ribbon width as shown by the insert in Fig.~\ref{fig:RamanWidth}(a). Similar linear relations have been observed in graphene and ML MoS$_2$ NRs~\cite{bischoff2011raman,wu2016monolayer}. The zero intercept at finite ribbon width ($\approx$~10~nm) is an indication that the effective ribbon width is narrower than the measured width, perhaps due to over etching~\cite{ryu2011raman, Wei2017}.

In graphene NRs and TMD ribbons prepared with other processes, defect modes and linewidth broadening are observed~\cite{ryu2011raman,wu2016monolayer}.
The Raman spectra of NRs measured here, however, exhibit similar lineshape to an unpatterned ML even in the narrowest NRs measured (Fig.~\ref{fig:RamanWidth}(b)). Our calculation (see supplementary materials for details) shows that linewidth broadening for Raman modes in monolayer MoS$_{2}$ NRs are within 1~cm$^{-1}$ for ribbons wider than 5~nm, which is beyond the resolution of a typical Raman system. In addition, no Raman modes on the low frequency side are present, indicating that the nanolithography process does not create appreciable structural damage to the ribbons and no edge modes have become measurable. This suggests that our approach represents a nondestructive fabrication process to the patterned ML optical properties, in contrast to recent ion milling approaches~\cite{wu2016monolayer}.  This conclusion is supported by electrical transport measurements showing comparable mobility in NRs of 40~nm width to that of a ML, except for edge effects which manifest in temperature-dependent mobility measurement (See supplementary materials for details).

\section{Photoluminescence Spectroscopy of MoS$_2$ Nanoribbons}

In nanomaterials and nanostructures, electron and hole wavefunctions are confined to small scales, resulting in a modification to their energy. Size-tunable optical bandgaps have been observed in graphene NRs due to the confinement effect~\cite{han2007energy}, and recently in ML MoS$_2$~\cite{Gopalakrishnan2015, Gan2015, Jin2016, Wei2017}. Similar confinement effects are expected in MoS$_{2}$ NRs, but they have not yet been reported~\cite{wu2016monolayer,poh2017large}. Our nondestructive fabrication process allows the study of PL from laterally-confined ML MoS$_2$ NRs.

\begin{figure}[tp]
\includegraphics[scale=1]{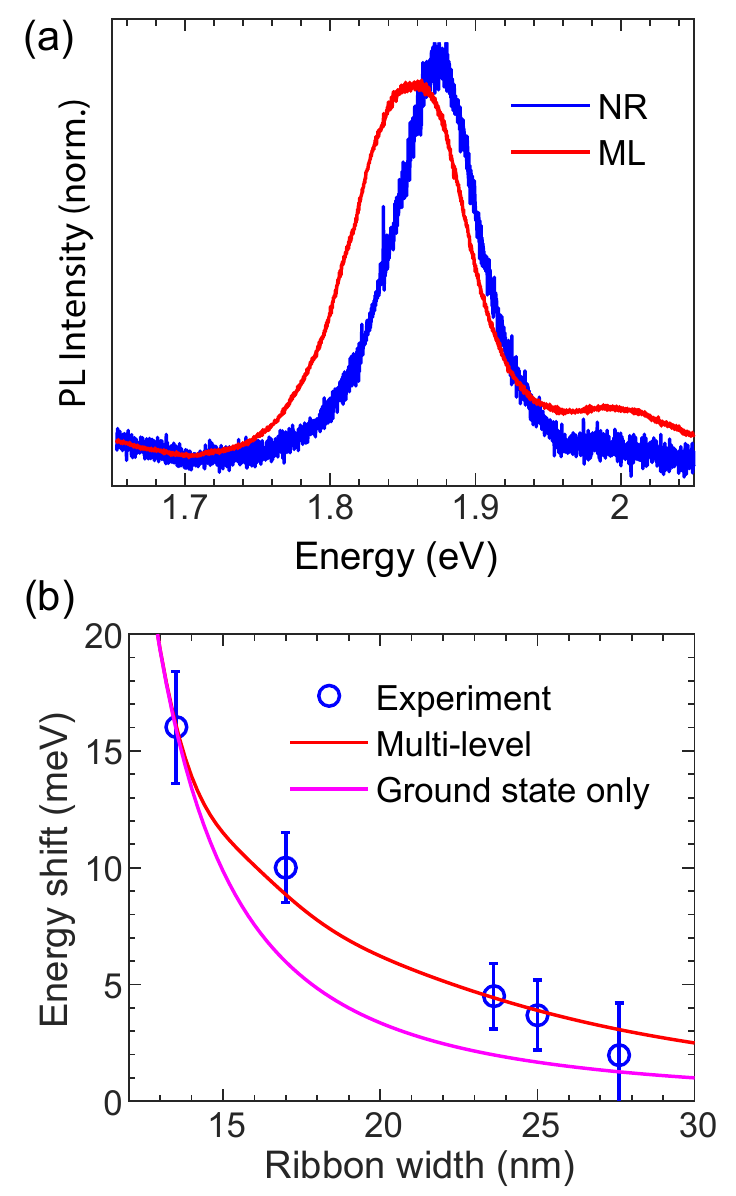}
\caption[PL spectrum of ML MoS$_{2}$ NRs]{PL spectra of ML MoS$_{2}$ and NRs. (a) PL spectra of ML (red) MoS$_{2}$ and NRs (blue). (b) PL peak energy shift with ribbon width. The red line is the fit to a multilevel model and the purple line is the fit with only the ground state been considered.}
\label{fig:PLWidth}
\end{figure}

Figure~\ref{fig:PLWidth}(a) shows the PL spectra of an unpatterned ML and NRs fabricated with the same flake. The NRs are measured with a nominal ribbon width of 14~nm. A clear blue-shifted PL peak is present, indicating the confinement effect due to the nanometer-scale lateral width. The spectral linewidth of NRs is slightly narrower (up to 10~meV narrowing), similar to the narrowing observed in ML MoS$_2$ nanodots~\cite{Wei2017}. This is possibly due to reduced phonon scattering in the confined system. Exfoliated ML MoS$_2$ generally exhibits $n$-doped properties with trion emission in PL~\cite{Mak2013,Zhang2014}, and defects along the edges should be considered as possible origins of the energy shift. However, defect induced trion emission in NRs can be excluded since trions possess lower energies than excitons, in contrast to the blue shifting observed here. Furthermore, the energy shift is size-dependent as shown in Fig.~\ref{fig:PLWidth} (b). The clear trend of larger shift in narrower ribbons suggests a confinement effect.

Since the Bohr radius of excitons in ML MoS$_{2}$ (and similar TMDs) is extremely small ($\sim1$~nm~\cite{ramasubramaniam2012large,cheiwchanchamnangij2012quasiparticle}), one can treat the excitons as quasiparticles whose center-of-mass (CM) motion is quantized in an infinite-deep potential well (Fig.~S4). This simple model gives the shift for the ground state as a function of ribbon width as
\begin{equation}
\Delta E=\frac{\hbar^2\pi^2}{2M_\textrm{\rm CM}W_\textrm{eff}^2}
\label{eq:dEribbon}
\end{equation}
where $M_\textrm{CM}$ is the exciton total mass and $W_\textrm{eff}$ is the effective ribbon width (Fig.~S4(a)). Similar to that observed in graphene NRs~\cite{han2007energy} and nanoflake quantum dots of MoS$_{2}$ MLs~\cite{Wei2017}, the effective ribbon width is typically several nanometers narrower than that measured with AFM (or scanning electron microscopy) due to the finite AFM tip size and over etching,  and the reconfigurable etched edges can also contribute to an narrower effective ribbon width through chemical reactions with water and oxygen in air~\cite{han2007energy,ryu2011raman}.  The experimental data is thus fit with $W_\textrm{eff}= W-\delta W$, where $W$ is the AFM measured nominal ribbon width. The fitting with the phenomenological Eq.~\ref{eq:dEribbon} gives $M_{\rm CM}=0.21 m_0$ and $\delta W = 3$~nm, where $m_0$ is the free electron mass. The fitted $M_{CM}=0.21 m_0$ is smaller than the theoretical values which are around $0.9m_0$, however, the simple model captures the $1/W^2 $ trend of the confinement energy. These size-dependent optical results in nanoribbons are very similar qualitatively to those reported for MoS$_2$ nanodots~\cite{Wei2017}.

In the weak confinement regime, the narrow energy level spacing and extremely short exciton lifetime in MoS$_2$ complicates emission dynamics from optically-excited excitons. Upon photo-excitation, hot excitons typically undergo a thermalization process through exciton-phonon and exciton-exciton interaction before recombination. However, due to the short radiative lifetime of the tightly bound excitons in ML TMDs (here MoS$_{2}$) and the reduced scattering rate with phonons in confined systems, excitons can recombine before reaching thermal equilibrium~\cite{robert2016exciton}. It is thus reasonable to consider the emissions from higher energy states since the exciton wavefunctions do not spatially redistribute. A similar phenomenon was also reported in conventional semiconductor systems in the weak confinement regime where high energy states occupation are suggested~\cite{Lee2011}.

To test this idea, a simple model accounting for the first three levels gives fitting parameters of $\delta W = 8$~nm and $M_\textrm{\rm CM}=0.78 m_0$. Figure~\ref{fig:PLWidth}(b) compares this to the typical ground state. The non-equilibrium multilevel model gives $M_{\rm CM}$ very close to the value of around $0.9 m_0$ reported in the literature~\cite{cheiwchanchamnangij2012quasiparticle,ramasubramaniam2012large}.
The effective 8~nm width reduction of $\delta W$ is comparable to the value observed for graphene NRs prepared with similar lithographically process~\cite{ryu2011raman}. It is also consistent with the 10~nm intercept from size-dependent Raman intensities in the inset in Fig.~\ref{fig:RamanWidth}(a). The non-equilibrium multilevel model for emission gives reasonable parameters compared to experiments for size-dependent features. We emphasize that this model is a simple approach to account for the size-dependence expected for fast recombining excitons in the weak confinement regime. A full dynamical model for the NR emission is beyond the scope of these experiments and analysis. Achieving strong lateral confinement with significant energy level separation and increased importance of edge states would lead to distinct PL behavior that has not yet been observed in ML TMD NRs.

\section{Conclusion}
We have shown that high resolution e-beam lithography can be used to pattern ML TMD nanoribbons with deterministic control of lateral size. The NRs prepared by e-beam lithography with positive resist do not show the large defect contributions to optical spectroscopy that appear in NRs prepared by ion milling. The isotropic Raman spectra in ML become anisotropic in NRs and the PL of the nanometer-scale ribbons shows confinement effects that can be modeled assuming a non-equilibrium system. The e-beam lithography approach offers a potential route for fabricating TMD NRs with exotic electrical and magnetic properties, particularly at the edges, that maintain good opto-electronic properties.

\section*{Acknowledgements\vspace{-.25em}}

This work is primarily supported by the Office of Naval research (N00014-16-1-3055) and the Institute for Sustainability and Energy at Northwestern. Characterization and device fabrication made use of the EPIC and SPID facilities of the NUANCE Center at Northwestern University and the Northwestern University Micro/ Nano Fabrication Facility (NUFAB), which have received support from the Soft and Hybrid Nanotechnology Experimental (SHyNE) Resource (NSF NNCI-1542205); the MRSEC program (NSF DMR-1121262) at the Materials Research Center; the International Institute for Nanotechnology (IIN); the Keck Foundation; and the State of Illinois.  Use of the Center for Nanoscale Materials was supported by the US Department of Energy, Office of Science, Office of Basic Energy Sciences, under Contract No. DE-AC02-06CH11357.  N.P.S. gratefully acknowledges support as an Alfred P. Sloan Research Fellow.

\section*{Supplementary Information}
Fabrication, size characterization, and modeling details are in the Supplementary Information file.

\bibliographystyle{apsrev4-1-title}
\bibliography{Wei2017}

\end{document}